# Au$_{13}$(8$e$): A Secondary Block for Describing a Special Group of Liganded Gold Clusters Containing Icosahedral Au$_{13}$ Motifs

Wen Wu Xu[1, 2], Xiao Cheng Zeng*[2,3], Yi Gao*[1, 4]

[1]*Division of Interfacial Water and Key Laboratory of Interfacial Physics and Technology, Shanghai Institute of Applied Physics, Chinese Academy of Sciences, Shanghai 201800, China.*
[2]*Department of Chemistry, University of Nebraska-Lincoln, Lincoln, Nebraska 68588, USA.*
[3]*Collaborative Innovation Center of Chemistry for Energy Materials, University of Science and Technology of China, Hefei, Anhui 230026, China*
[4]*Shanghai Science Research Center, Chinese Academy of Sciences, Shanghai 201204, China*

*Correspondence Email: xzeng1@unl.edu, gaoyi@sinap.ac.cn*

**Abstract**

A grand unified model (GUM) has been proposed recently to understand structure anatomy and evolution of liganded gold clusters. In this work, besidesthe two types of elementary blocks (triangular Au$_3$(2e) and tetrahedral Au$_4$(2e)), we introduce a secondary block, namely, the icosahedral Au$_{13}$ with 8e valence electrons, noted as Au$_{13}$(8e). Using this secondary block, structural anatomy and evolution of a special group of liganded gold nanoclusters containing icosahedral Au$_{13}$ motifs can be conveniently analyzed. In addition, a new ligand-protected cluster Au$_{49}$(PR$_3$)$_{10}$(SR)$_{15}$Cl$_2$ is predicted to exhibit high chemical and thermal stability, suggesting likelihood of its synthesis in the laboratory.

**1. Introduction**

Ligand-protected gold (Au) clusters have attracted considerable interests due to their broad applications in electrochemistry, nanocatalysis and bioengineering.[1-4] From 1970s to the present, a large number of liganded gold nanoclusters have been synthesized and their structures have been determined via X-ray crystallography [5-10] or predicted by density functional theory (DFT) computation.[11-17] Although total structural determination and prediction of numerous ligand-protected gold nanoclusters have been accomplished, understanding of structural stabilities and evolution of many seemingly-unrelated structures of the liganded gold nanoclusters remains to be an active area of research. In 1970s, the Wade-Mingos electron-counting rule was proposed and later employed in the cluster chemistry.[18,19] However, this rule of thumb cannot fully describe ligand-protected Au nanoclusters.[20] In 2008, Walter and Häkkinen *et al.* proposed the superatom complex (SAC) model to explain high stabilities of some spherical ligand-protected

Au nanoclusters with total count of valence electrons of 2, 8, 18, 34, 58 (electronic shell closing) on basis of the jellium model.[21] Later, Cheng *et al.* developed a super valence bond (SVB) model [22] and a superatom network (SAN) model [23] to interpret high stabilities of certain nonspherical thiolate-protected Au nanoclusters based on the adaptive natural density partitioning (AdNDP) analysis.[24] A key concept in SAN model is that the core of Au nanoclusters can be viewed as a network of n-centered two-electron (n = 2–6) superatoms, originally developed to describe some small-sized Au clusters by Sergeeva and Boldyrev.[25] Mingos analyzed some ligand-protected Au nanoclusters containing vertex, edge, or face-sharing icosahedral or cuboctahedra $Au_{13}$ units, by using the notion of the united atom model for diatomic molecules (i.e., $F_2$, $O_2$, $N_2$ etc.).[26] Very recently, the Borromean-ring diagrams [27] were utilized by Pradeep, Whetten and coworkers to analyze high stabilities of $[Au_{25}(SR)_{18}]^{1-}$,[9-11] $Au_{38}(SR)_{24}$,[28] and $Au_{102}(SR)_{44}$.[6] Again, because of the wide variety of ligand-protected Au nanoclusters with different core morphologies, protection ligands, and number of valence electrons, none of these models can offer a unified description of their structure anatomy, structural stabilities and evolution.

Recently, a grand unified model (GUM) has been proposed to gain fundamental understanding of a plethora of complex and seemingly-unrelated structures of liganded gold clusters.[29] In GUM, depending on the connected chemical group of protection ligands, gold atoms on surface of the gold core can be assigned to three possible valence states (in names of three flavors), i.e., bottom (1*e*), middle (0.5*e*), and top (0*e*) flavor. The gold cores of all 71 ligand-protected clusters reported in the literature can be universally viewed as packing of two types of elementary blocks: triangular $Au_3$ and tetrahedral $Au_4$, each satisfying the duet rule [noted as $Au_3(2e)$ and $Au_4(2e)$ elementary blocks]. Such a generic rule of thumb can be used to characterize the stabilities and structural evolution of liganded gold clusters. Among the 71 ligand-protected clusters, there is a special group of liganded gold nanoclusters all containing one or several icosahedral $Au_{13}$ motifs, such as $[Au_{13}(PR_3)_{10}Cl_2]^{3+}$,[30] $[Au_{13}(dppe)_5Cl_2]^{3+}$,[31] $Au_{16}(AsPh_3)_8Cl_6$,[32] $[Au_{19}(C{\equiv}CR)_9(Hdppa)_3]^{2+}$,[33] $Au_{20}(PP_3)_4Cl_4$,[34] $[Au_{25}(PR_3)_{10}(SR)_5Cl_2]^{2+}$,[35] $[Au_{25}(SR)_{18}]^{1-}$,[9-11] $[Au_{37}(PR_3)_{10}(SR)_{10}Cl_2]^{1+}$,[36] and $Au_{38}(SR)_{24}$.[28,37] For these liganded clusters, according to the electron counting protocols for effective detachment of ligands in GUM, each icosahedral $Au_{13}$ motif is assigned to have 8*e* valence electrons [$Au_{13}(8e)$]. In fact, each icosahedral $Au_{13}$ motif can be viewed as packing of four elementary blocks. For example, the $Au_{13}(8e)$ in $[Au_{25}(SR)_{18}]^{1-}$ consists of two elementary blocks of $Au_3(2e)$ and two $Au_4(2e)$, as shown in Figure S1. The AdNDP analysis (Figure S2) also confirms that the $Au_{13}(8e)$ can be decomposed into four elementary blocks. Note that this decomposition is not intended to reflect the electronic structure of the $Au_{13}$ core at the molecular level (Figure S3) but simply to indicate that the 8*e* valence electrons of $Au_{13}(8e)$ can be viewed as a sum of four pairs of valence electrons of the four elementary blocks. The elementary blocks $Au_3(2e)$ and $Au_4(2e)$ can be also viewed as electron shell closure species, in analogue of that of the stable He atom.

Likewise, $Au_{13}(8e)$ can be viewed as an electron shell closure species, in analogue of that of the stable Ne atom. As such, the $Au_{13}(8e)$ may be regarded as a secondary block (or coarse-grained block) to constitute the gold cores of a special group of liganded gold clusters. As such, 91 variants of valence states (named as $I_1 - I_{91}$) for the secondary block $Au_{13}(8e)$ can be identified (see Table S1). In addition, the structural evolution of $[Au_{13}(PR_3)_{10}Cl_2]^{3+}$, $[Au_{25}(PR_3)_{10}(SR)_5Cl_2]^{2+}$, and $[Au_{37}(PR_3)_{10}(SR)_{10}Cl_2]^{1+}$ can be analyzed more conveniently on the basis of the secondary block $Au_{13}(8e)$. Furthermore, a new cluster $Au_{49}(PR_3)_{10}(SR)_{15}Cl_2$ is predicted based on the structural trend obtained with the three experimentally synthesized clusters.

## 2. Computational methods

Structural optimizations of the liganded gold clusters were performed using density functional theory (DFT) methods implemented in the Gaussian 09 program package [38] The M06L functional [39] and the all-electron basis set 6-31G* for H, P, S, and Cl, effective-core basis set LANL2DZ for Au were selected. *Ab initio* molecular dynamics simulation was performed in the canonical ensemble at a finite temperature of 355 K using DFT method implemented in the CP2K package.[40,41] The Nóse-Hoover thermostat and a time step of 1 fs were employed. The exchange correlation potential was described by the generalized-gradient approximation (GGA) with the spin-polarized functional of Perdew-Burke-Ernzerh (PBE).[42] Wavefunctions were expanded in triple-ζ Gaussian basis sets with an auxiliary plane-wave basis and a cutoff energy of 300 Ry. Core electrons were treated by scalar relativistic norm-conserving pseudopotentials with 11, 7, 6, 5, and 1 valence electrons for Au, Cl, S, P, and H, respectively.[43,44] Brillouin zone integration was undertaken with using a reciprocal-space mesh consisting of only the Γ-point. The –R group is replaced by –H to significantly lower computational cost.

## 3. Results and discussions

In the ligand-protected gold nanoclusters, there are several types of protected ligands such as bridged –SR– group, gold–thiolate "staple motifs", halide X (X = F, Cl, Br, and I), and $PR_3/AsR_3$ functional groups. In order to isolate the gold cores of ligand-protected gold nanoclusters, the protected ligands should be effectively detached from the gold cores. The following electron-counting protocols for effective separation of different types of protected ligands from the gold cores for the ligand-protected gold nanoclusters can be undertaken.

(a) When two gold atoms bonded with a bridged –SR– group (Figure S4a) or a gold–thiolate "staple motif" (Figure S4b), each gold atom should transfer $0.5e$ valence electron to bridged –SR– group or gold–thiolate "staple motif" so that the bridged –SR– group and gold–thiolate "staple motif" can be detached. Therefore, both of gold atoms are assigned to have $0.5e$ valence electron and the middle flavor.

(b) Gold atom bonded with X (X = F, Cl, Br, and I) transfers 1*e* valence electron to X to be detached (Figure S4c). Therefore, the gold atom is assigned to have 0*e* valence electron and the top flavor.

(c) For gold atom bonded with $PR_3$/$AsR_3$ functional groups, the $PR_3$/$AsR_3$ functional groups can be detached directly without needing to transfer the valence electrons (Figure S4d), resulting in the 1*e* valence electron and bottom flavor assigned to the associated gold atom.

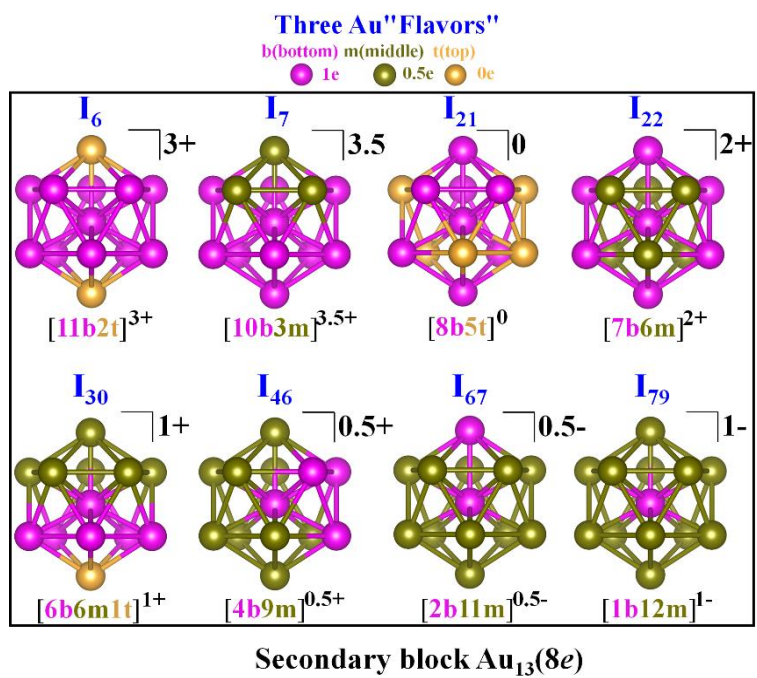

**Figure 1.** 8 of the 91 variants ($I_1 – I_{91}$) of valence states for the secondary block $Au_{13}(8e)$ in $[Au_{13}(PR_3)_{10}Cl_2]^{3+}$, $[Au_{13}(dppe)_5Cl_2]^{3+}$, $Au_{16}(AsPh_3)_8Cl_6$, $[Au_{19}(C≡CR)_9(Hdppa)_3]^{2+}$, $Au_{20}(PP_3)_4Cl_4$, $[Au_{25}(PR_3)_{10}(SR)_5Cl_2]^{2+}$, $[Au_{25}(SR)_{18}]^{1-}$,[9-11] $[Au_{37}(PR_3)_{10}(SR)_{10}Cl_2]^{1+}$, and $Au_{38}(SR)_{24}$ clusters. b, m, and t denote bottom flavor (1e), middle flavor (0.5e), and top flavor (0e), respectively. Color code: Au – magenta ("bottom flavor"), dark yellow ("middle flavor"), and yellow ("top flavor").

It should be noted that different protection ligands can result in three different valence states of surface gold atoms, i.e., 1*e*, 0.5*e*, and 0*e*, which correspond to the three flavors, i.e., bottom (1*e*), middle (0.5*e*), and top (0*e*). Therefore, the secondary block $Au_{13}(8e)$ has 91 variants of valence states (named as $I_1 – I_{91}$), as presented in Table S2. The representative valence states for the secondary blocks of $Au_{13}(8e)$ in $[Au_{13}(PR_3)_{10}Cl_2]^{3+}$,[30] $[Au_{13}(dppe)_5Cl_2]^{3+}$,[31] $Au_{16}(AsPh_3)_8Cl_6$,[32] $[Au_{19}(C≡CR)_9(Hdppa)_3]^{2+}$,[33] $Au_{20}(PP_3)_4Cl_4$,[34] $[Au_{25}(PR_3)_{10}(SR)_5Cl_2]^{2+}$,[35] $[Au_{25}(SR)_{18}]^{1-}$,[9-11] $[Au_{37}(PR_3)_{10}(SR)_{10}Cl_2]^{1+}$,[36] and $Au_{38}(SR)_{24}$ [28,37] clusters are shown in Figure 1, in which b in magenta, m in dark yellow, and t in yellow denote bottom flavor (1*e*), middle flavor (0.5*e*), and top flavor (0*e*), respectively. For example, with using the three Au "flavors" b, m, and t, $I_{79}$ can be rewritten as [1b12m]$^{1-}$. The total valence electrons of I$_{79}$ can be computed as $1 \times 1e + 12 \times 0.5e + 1e = 8e$ (Figure S5).

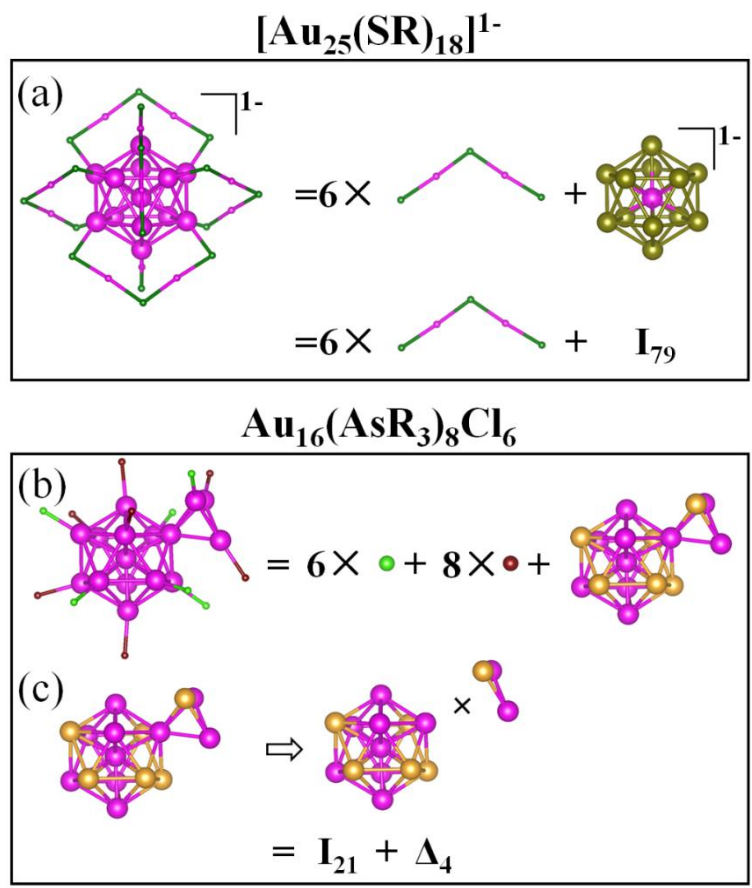

**Figure 2.** Structural decompositions of [Au$_{25}$(SR)$_{18}$]$^{1-}$ (a) and Au$_{16}$(AsPh$_3$)$_8$Cl$_6$ (b and c). Au – magenta ("bottom flavor"), dark yellow ("middle flavor"), and yellow ("top flavor"); SR – dark green; Cl – light green; AsR$_3$ – wine. The R groups are omitted for clarity.

Next, four prototype structures are used as two examples to analyze the valence state of the secondary block Au$_{13}$(8$e$). The first is [Au$_{25}$(SR)$_{18}$]$^{1-}$,[9-11] a cluster widely investigated experimentally and theoretically.[45,46] As shown in Figure 2a, the icosahedral Au$_{13}$ core of [Au$_{25}$(SR)$_{18}$]$^{1-}$ is protected by six [SR-Au-SR-Au-RS] staple motifs. According to electron counting protocols of the GUM, each of the two surface gold atoms bonded with SR is assigned to transfer 0.5$e$ valence electron to the staple motif.[29] Thus, each of the 12 surface gold atoms of the icosahedral Au$_{13}$ core is assigned to have 0.5$e$ valence electron. Since the central gold atom of the icosahedral Au$_{13}$ core has 1$e$ valence electron and the net charge of [Au$_{25}$(SR)$_{18}$]$^{1-}$ is -1, the total valence electrons of the icosahedral Au$_{13}$ core is $12 \times 0.5e + 1e + 1e = 8e$, and the icosahedral Au$_{13}$ core is in the I$_{79}$ valence state.

The second prototype structure considered is Au$_{16}$(AsPh$_3$)$_8$Cl$_6$.[32] It can be seen from Figure 2b and Figure 2c that the Au$_{16}$ core Au$_{16}$(AsPh$_3$)$_8$Cl$_6$ can be viewed as an icosahedral $Au_{13}$ motif combined with a triangular $Au_3$ motif. Each gold atom bonded with Cl/$AsR_3$ group is assigned to have $1e/0e$ valence electron. The total number of valence electrons of the icosahedral $Au_{13}$ motif and triangular $Au_3$ are $10e$. Thus, the $Au_{16}$ core in $Au_{16}(AsPh_3)_8Cl_6$ can be decomposed as one secondary block $Au_{13}(8e)$ in the $I_{21}$ valence state, and one elementary block $Au_3(2e)$ in the $\Delta_4$ valence state.

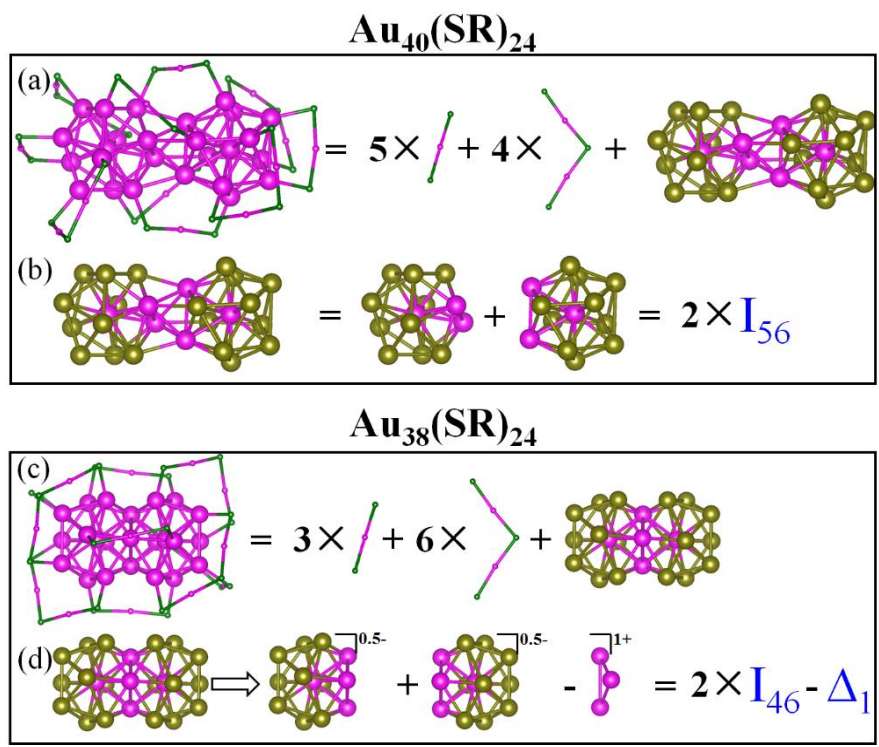

**Figure 3.** Structural decompositions of $Au_{40}(SR)_{24}$ (a and b) and $Au_{38}(SR)_{24}$ (c and d). Au – magenta ("bottom flavor") and dark yellow ("middle flavor"); SR – dark green. The R groups are omitted for clarity.

The third prototype structure considered is a theoretical structure $Au_{40}(SR)_{24}$. As shown in Figure 3a, $Au_{40}(SR)_{24}$ has a bi-icosahedral $Au_{26}$ core protected by five [RS–Au–SR] staples and four [RS–Au–SR–Au–SR] staples. Then the $Au_{26}$ core with $16e$ valence electrons can be decomposed into two $I_{56}$ secondary blocks packed together (Figure 3b). In Figure 3c, the fourth prototype structure $Au_{38}(SR)_{24}$ has a face-fused bi-icosahedral $Au_{23}$ core protected by three [RS–Au–SR] staples and six [RS–Au–SR–Au–SR] staples. According to the secondary elementary block $Au_{13}(8e)$, the $Au_{23}$ core, as shown in Figure 4b, can be viewed as two $I_{46}$ secondary blocks fused together by sharing one $\Delta_1$ elementary block (Figure 3d). This is very different from previous structural analyses on $Au_{38}(SR)_{24}$ using SVB model, where two 13-centered 7-electron icosahedral units were proposed. In addition, the $Au_{20}$ core of $Au_{20}(PP_3)_4^{4+}$ can be divided into five blocks $I_7$, $T_5$, and 3 $\Delta_4$, as shown in Figure S6.

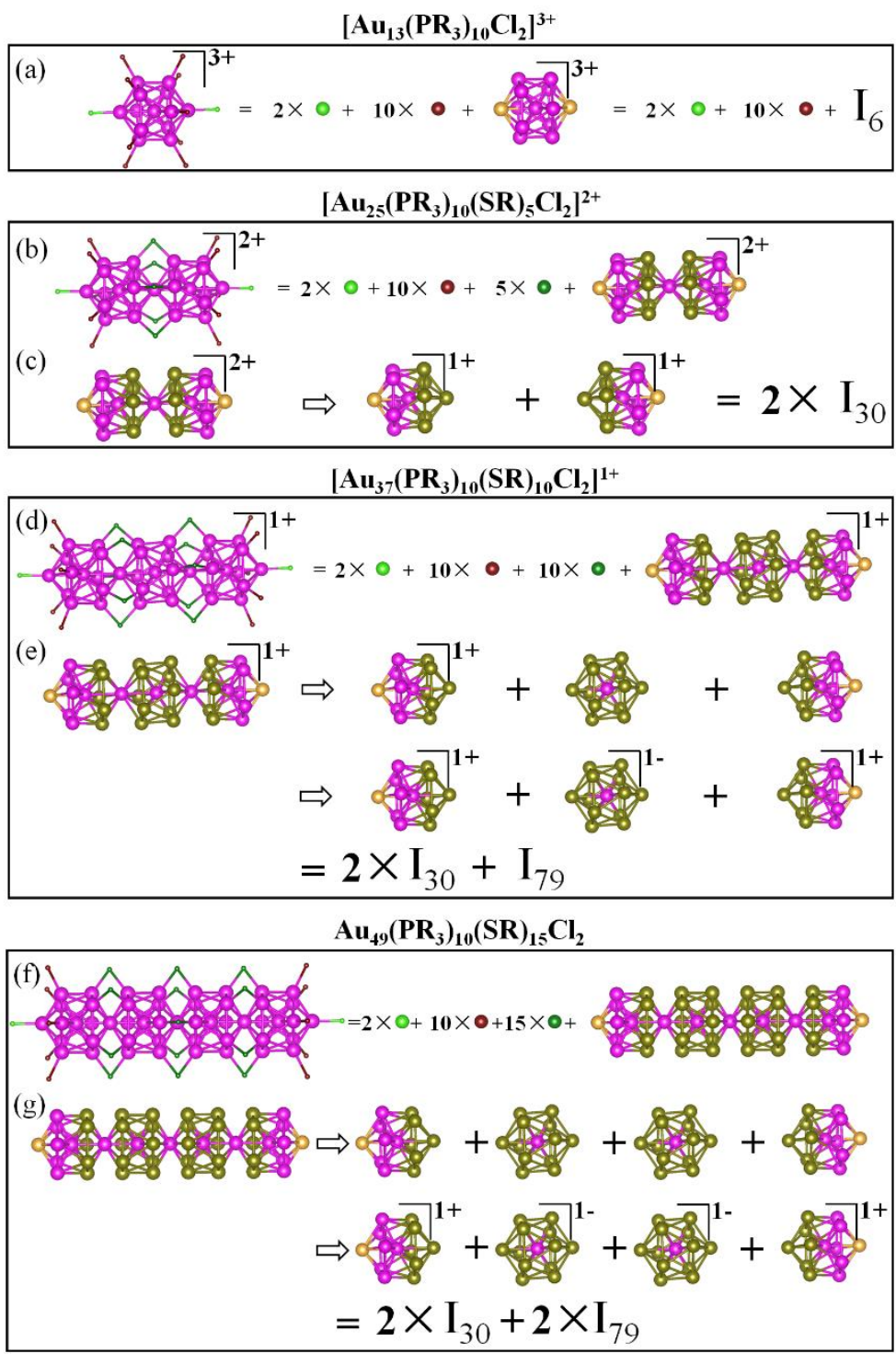

Figure 4. A structural decomposition of $[Au_{13}(PR_3)_{10}Cl_2]^{3+}$ (a), $[Au_{25}(PR_3)_{10}(SR)_5Cl_2]^{2+}$ (b and c), $[Au_{37}(PR_3)_{10}(SR)_{10}Cl_2]^{1+}$ (d and e), and $Au_{49}(PR_3)_{10}(SR)_{15}Cl_2$ (f and g). Color code: Au – magenta ("bottom flavor"), dark yellow ("middle flavor"), and yellow ("top flavor"); SR – dark green; Cl – light green; $PR_3$ – wine. The R groups are omitted for clarity.

In Figure 4, structural decompositions of $[Au_{13}(PR_3)_{10}Cl_2]^{3+}$,[30] $[Au_{25}(PR_3)_{10}(SR)_5Cl_2]^{2+}$,[35] and $[Au_{37}(PR_3)_{10}(SR)_{10}Cl_2]^{1+}$ [36] are displayed. The details of counting the valence electrons of the $Au_{13}$ unit in these structures can refer to Figures S7-S10. It can be seen that the icosahedral $Au_{13}$ core of $[Au_{13}(PR_3)_{10}Cl_2]^{3+}$ is in the $I_6$ valence state (Figure 4a). The $Au_{25}$ core in the $[Au_{25}(PR_3)_{10}(SR)_5Cl_2]^{2+}$ cluster can be viewed as two icosahedral $Au_{13}$ motifs fused together by sharing one Au atom (Figure 4b). Each icosahedral $Au_{13}$ motif, with 8 valence electrons, is in the $I_{30}$ valence state (Figure 4c). Likewise, the $Au_{37}$ core of $[Au_{37}(PR_3)_{10}(SR)_{10}Cl_2]^{1+}$ cluster can be viewed as three icosahedral $Au_{13}$ motifs fused together by sharing two Au atoms (Figure 4d). Specifically, the three secondary blocks $Au_{13}(8e)$ are in the $I_{30}$, $I_{30}$, and $I_{79}$ valence state, respectively, giving rise to totally 24 valence electrons for the core of $[Au_{37}(PR_3)_{10}(SR)_{10}Cl_2]^{1+}$ cluster (Figure 4e).

Table 1. Computed properties of $[Au_{13}(PR_3)_{10}Cl_2]^{3+}$, $[Au_{25}(PR_3)_{10}(SR)_5Cl_2]^{2+}$, $[Au_{37}(PR_3)_{10}(SR)_{10}Cl_2]^{1+}$, and $Au_{49}(PR_3)_{10}(SR)_{15}Cl_2$. ∗ denotes a predicted structure based on GUM.

|  | Charge | Number of valence electrons | Number of $Au_{13}(8e)$ | HOMO-LUMO gap/eV |
| --- | --- | --- | --- | --- |
| $[Au_{13}(PR_3)_{10}Cl_2]^{3+}$ | +3 | 8 | 1 | 1.96 |
| $[Au_{25}(PR_3)_{10}(SR)_5Cl_2]^{2+}$ | +2 | 16 | 2 | 1.31 |
| $[Au_{37}(PR_3)_{10}(SR)_{10}Cl_2]^{1+}$ | +1 | 24 | 3 | 0.90 |
| $Au_{49}(PR_3)_{10}(SR)_{15}Cl_2$ ∗ | 0 | 32 | 4 | 0.79 |

It should be noted that the charges of $[Au_{13}(PR_3)_{10}Cl_2]^{3+}$, $[Au_{25}(PR_3)_{10}(SR)_5Cl_2]^{2+}$, and $[Au_{37}(PR_3)_{10}(SR)_{10}Cl_2]^{1+}$ decrease from +3 to +1 with increasing the number of the secondary blocks of $Au_{13}(8e)$, as shown in Table 1. This trend suggests that the adjustment of the overall charge state is correlated with allowing each icosahedral $Au_{13}$ motif to have 8 valence electrons. If this empirical trend can be extended to a larger-sized cluster whose gold core has four secondary blocks of $Au_{13}(8e)$, a neutral $Au_{49}(PR_3)_{10}(SR)_{15}Cl_2$ cluster with 32 valence electrons can be predicted (see Table 1 and Figure 3f). Here, the four secondary blocks $Au_{13}(8e)$ are in the $I_{30}$, $I_{30}$, $I_{79}$, and $I_{79}$ valence state (Figure 3g), respectively. This newly predicted $Au_{49}(PR_3)_{10}(SR)_{15}Cl_2$ cluster exhibits a computed HOMO-LUMO gap of 0.79 eV, suggesting that this cluster is likely chemically stable. On the other hand, the HOMO-LUMO gaps for this series of clusters suggest that the chemical stabilities from $[Au_{13}(PR_3)_{10}Cl_2]^{3+}$ to $Au_{49}(PR_3)_{10}(SR)_{15}Cl_2$ may become weaker with increasing the number of secondary blocks $Au_{13}(8e)$. The 10 ps *ab initio* molecular dynamics (AIMD) simulation performed for the $Au_{49}(PR_3)_{10}(SR)_{15}Cl_2$ cluster suggested that the cluster may be thermally stable at 355 K (Figure 5). Lastly, the HOMOs and LUMOs of

$[Au_{13}(PR_3)_{10}Cl_2]^{3+}$, $[Au_{25}(PR_3)_{10}(SR)_5Cl_2]^{2+}$, $[Au_{37}(PR_3)_{10}(SR)_{10}Cl_2]^{1+}$, and $Au_{49}(PR_3)_{10}(SR)_{15}Cl_2$ are presentedn in Figure S11, from which the structural evolution of the secondary blocks $Au_{13}(8e)$ can be seen.

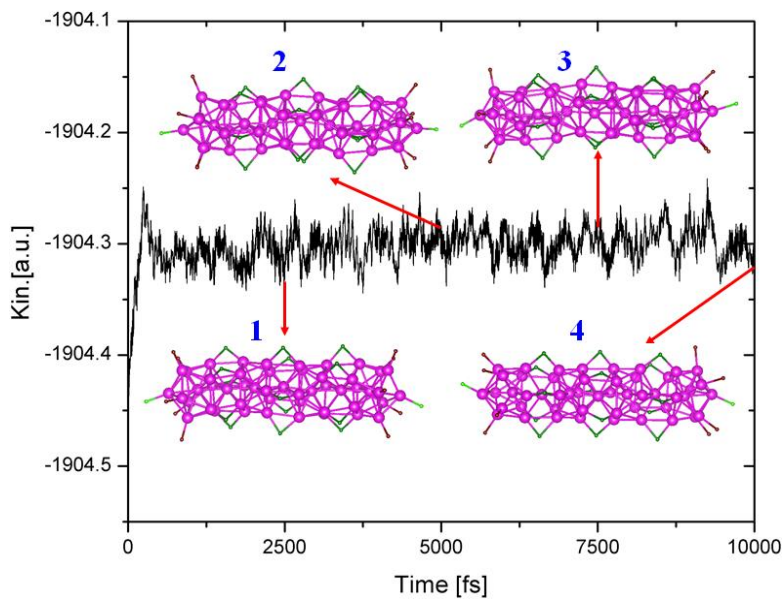

Figure 5. Computed kinetic energy (Kin) vs simulation time in an *ab initial* molecular dynamics simulation of the $Au_{49}(PH_3)_{10}(SH)_{15}Cl_2$ cluster at 355 K. 1, 2, 3, and 4 correspond to the snapshot at 2.5, 5, 7.5, and 10 ps, respectively. Color code: Au – magenta; SR – dark green; $PR_3$ – wine. The H atoms are omitted for clarity.

## 4 Conclusions

We have proposed that the $Au_{13}$ with 8 valence electrons can be conveniently viewed as a secondary block to describe structure anatomy and structural evolution of a special group of liganded gold clusters containing single or multiple icosahedral $Au_{13}$ motifs. Such a secondary block, noted as $Au_{13}(8e)$, can exhibit 91 variants of valence states ($I_1 - I_{91}$) due to three possible "flavors" [bottom (1*e*), middle (0.5*e*), and top (0*e*)] of gold atoms. The structural evolution from $[Au_{13}(PR_3)_{10}Cl_2]^{3+}$ to $[Au_{25}(PR_3)_{10}(SR)_5Cl_2]^{2+}$ and $[Au_{37}(PR_3)_{10}(SR)_{10}Cl_2]^{1+}$ is illustrated with the increasing number of the secondary blocks. We found that the charges of $[Au_{13}(PR_3)_{10}Cl_2]^{3+}$, $[Au_{25}(PR_3)_{10}(SR)_5Cl_2]^{2+}$, and $[Au_{37}(PR_3)_{10}(SR)_{10}Cl_2]^{1+}$ decrease from +3 to +1 with the increase of the number of secondary blocks $Au_{13}(8e)$ from 1 to 3. Based on this trend, a new cluster, $Au_{49}(PR_3)_{10}(SR)_{15}Cl_2$ with four secondary blocks $Au_{13}(8e)$, is predicted. This cluster seems to have high chemical and thermal stability, thereby may be synthesized in the laboratory. The introduction of secondary block $Au_{13}(8e)$ into the GUM appears to be a convenient supplement to understand a special group of liganded gold clusters containing icosahedral $Au_{13}$ motifs, and thus may be exploited for predicting new liganded gold clusters by design.

**Acknowledgements**

W.W.X. is supported by National Natural Science Foundation of China (11504396). Y.G. is supported by the start-up funding from Shanghai Institute of Applied Physics, Chinese Academy of Sciences (Y290011011), National Natural Science Foundation of China (21273268, 11574340), 'Hundred People Project' from Chinese Academy of Sciences and CAS-Shanghai Science Research Center (CAS-SSRC-YJ-2015-01). X.C.Z. is supported by a grant from Nebraska Center for Energy Sciences Research and a Qian-ren B (One Thousand Talent Plan B) summer research fund from USTC, and by a State Key R&D Fund of China (2016YFA0200604) to USTC. The computational resources utilized in this research were provided by Shanghai Supercomputer Center, National Supercomputing Center in Tianjin and Shenzhen, special program for applied research on super computation of the NSFC-Guangdong joint fund (the second phase) and NC3 computer facility in University of Nebraska-Lincoln.

# Supplementary Information

**Au$_{13}$(8$e$): A Secondary Block for Describing a Special Group of Liganded Gold Clusters Containing Icosahedral Au$_{13}$ Motifs**

Wen Wu Xu[1,2], Xiao Cheng Zeng*[2,3], Yi Gao*[1,4]

[1]Division of Interfacial Water and Key Laboratory of Interfacial Physics and Technology, Shanghai Institute of Applied Physics, Chinese Academy of Sciences, Shanghai 201800, China.

[2]Department of Chemistry, University of Nebraska-Lincoln, Lincoln, Nebraska 68588, USA.

[3]Collaborative Innovation Center of Chemistry for Energy Materials, University of Science and Technology of China, Hefei, Anhui 230026, China

[4]Shanghai Science Research Center, Chinese Academy of Sciences, Shanghai 201204, China

*Correspondence Email: xzeng1@unl.edu, gaoyi@sinap.ac.cn

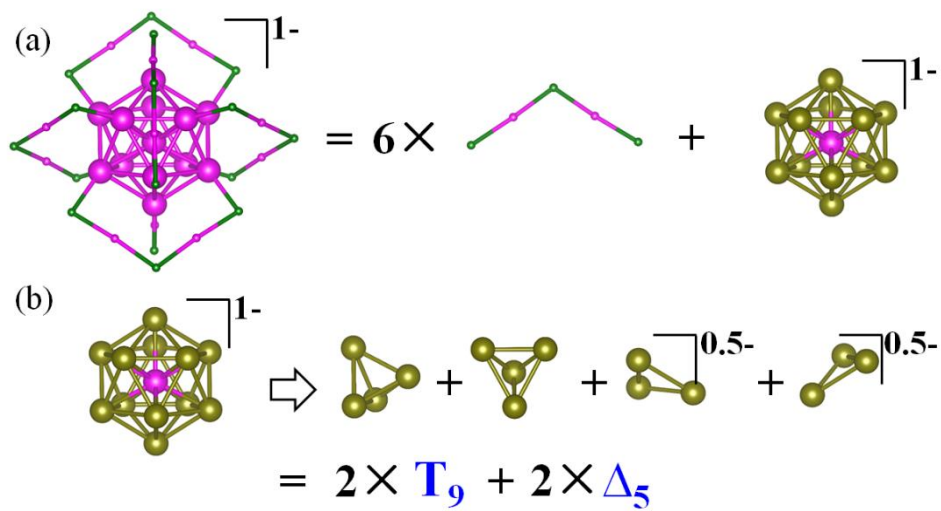

Figure S1. A structural decomposition of $[Au_{25}(SR)_{18}]^{1-}$ (a and b) into four elementary blocks. Au – magenta ("bottom flavor") and dark yellow ("middle flavor"); SR – dark green. The R groups are omitted for clarity.

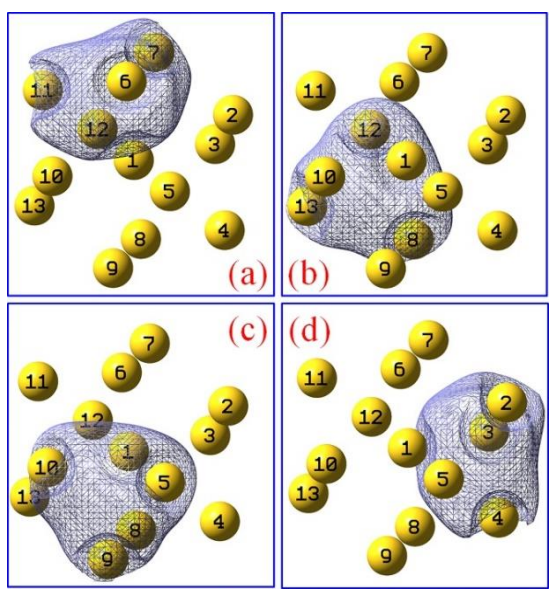

Figure S2. AdNDP analysis of the secondary block $Au_{13}(8e)$. The occupation number of the four elementary blocks (blue mesh) $Au_4(2e)$ in (a), (b), (c), and (d) are 1.63. The numbers 1-13 denote the 13 gold atoms. Color code: Au – yellow.

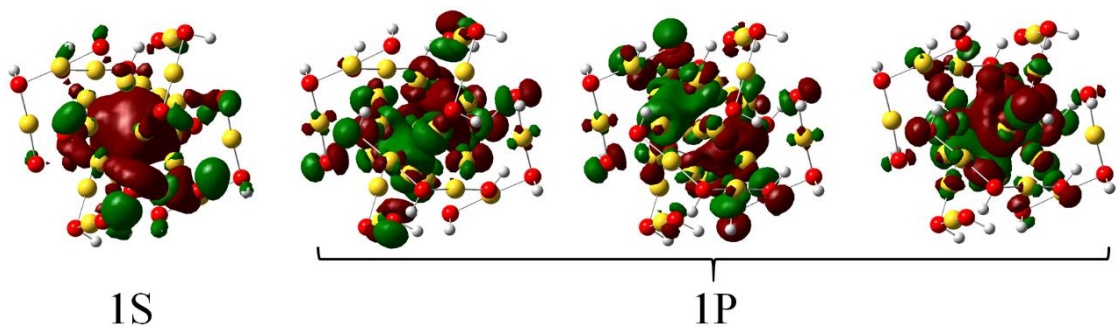

Figure S3. Computed delocalized occupied orbitals (1S and 1P) of $[Au_{25}(SR)_{18}]^{1-}$. Color code: Au – yellow; S – red; H – white.

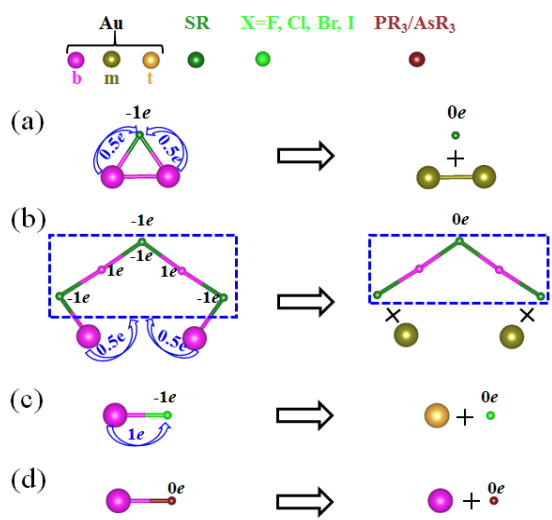

Figure S4. Electron counting protocols for effective detachment of ligands from inner Au core in the following four cases: Au atom bonded with (a) SR, (b) gold–thiolate "staple motifs", (c) X (X = F, Cl, Br, and I), and (d) PR3 functional groups. The blue arrow denotes charge transfer. Color code: Au – magenta ("bottom flavor"), dark yellow ("middle flavor"), and yellow ("top flavor"); S – dark green; X – light green; P - wine. The R groups are omitted for clarity.

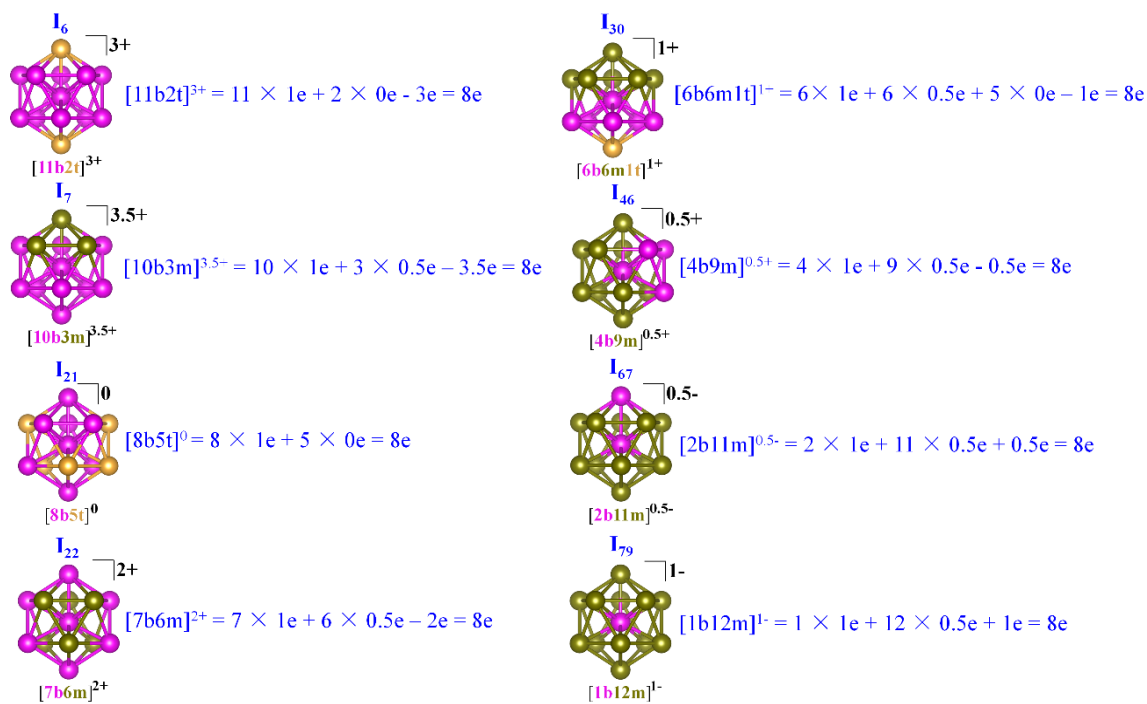

Figure S5. Counting the valence electrons of $Au_{13}$ clusters listed in Figure1. Color code: Au – magenta ("bottom flavor"), dark yellow ("middle flavor"), and yellow ("top flavor").

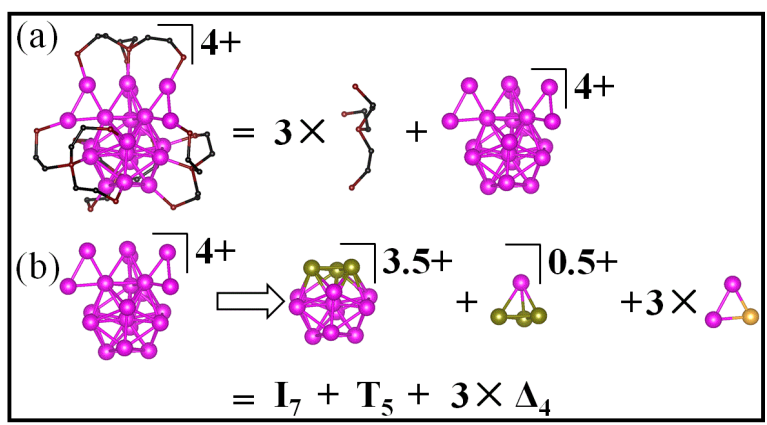

Figure S6. Structural decomposition of $[Au_{20}(PP_3)_4]^{4+}$ (a and b). Au – magenta ("bottom flavor"), dark yellow ("middle flavor"), and yellow ("top flavor"); C – black; $PR_3$ – wine. The R groups are omitted for clarity.

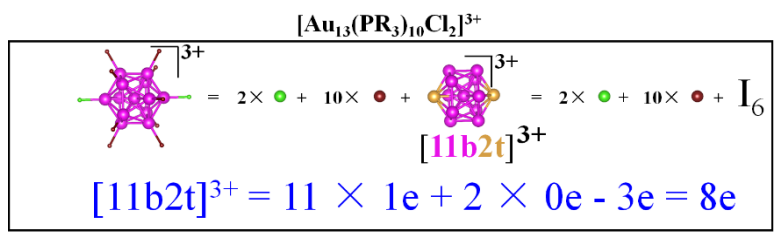

Figure S7. Counting the valence electrons of Au$_{13}$ core of [Au$_{13}$(PR$_3$)$_{10}$Cl$_2$]$^{3+}$. Color code: Au – magenta ("bottom flavor") and yellow ("top flavor"); S – dark green; X – light green; P - wine. The R groups are omitted for clarity.

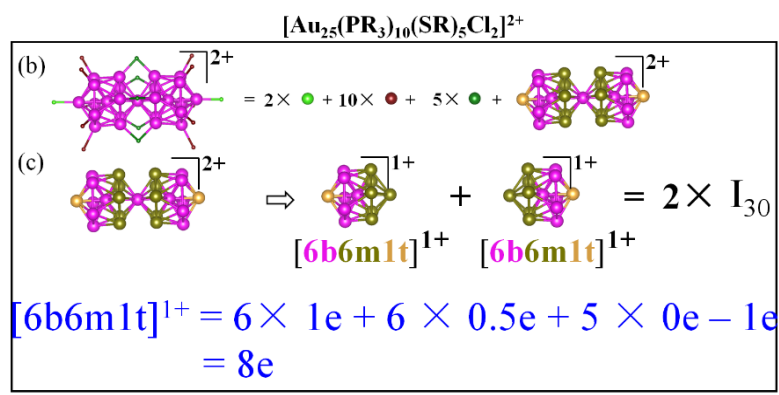

Figure S8. Counting the valence electrons of two Au$_{13}$ in Au$_{25}$ core of [Au$_{25}$(PR$_3$)$_{10}$(SR)$_5$Cl$_2$]$^{2+}$. Color code: Au – magenta ("bottom flavor"), dark yellow ("middle flavor"), and yellow ("top flavor"); S – dark green; X – light green; P - wine. The R groups are omitted for clarity.

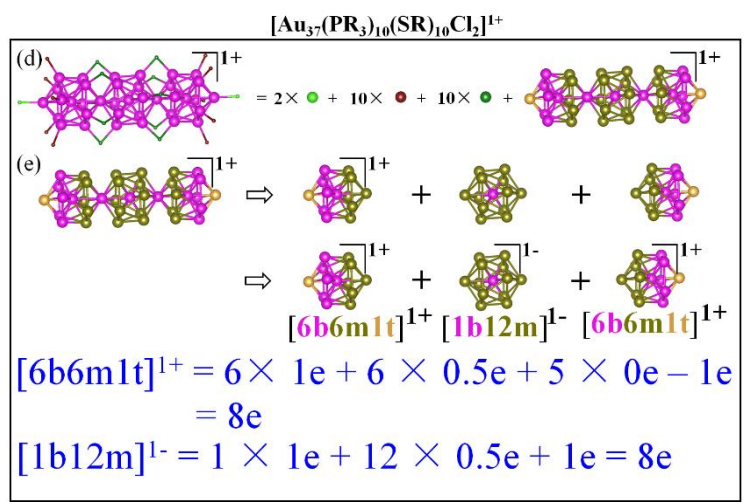

Figure S9. Counting the valence electrons of three Au$_{13}$ in Au$_{37}$ core of [Au$_{37}$(PR$_3$)$_{10}$(SR)$_{10}$Cl$_2$]$^{1+}$. Color code: Au – magenta ("bottom flavor"), dark yellow ("middle flavor"), and yellow ("top flavor"); S – dark green; X – light green; P - wine. The R groups are omitted for clarity.

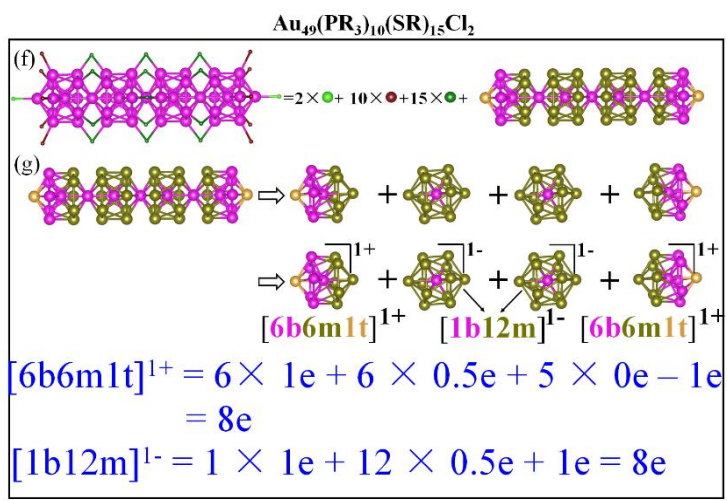

Figure S10. Counting the valence electrons of four Au$_{13}$ in the Au$_{49}$ core of Au$_{49}$(PR$_3$)$_{10}$(SR)$_{15}$Cl$_2$. Color code: Au – magenta ("bottom flavor"), dark yellow ("middle flavor"), and yellow ("top flavor"); S – dark green; X – light green; P - wine. The R groups are omitted for clarity.

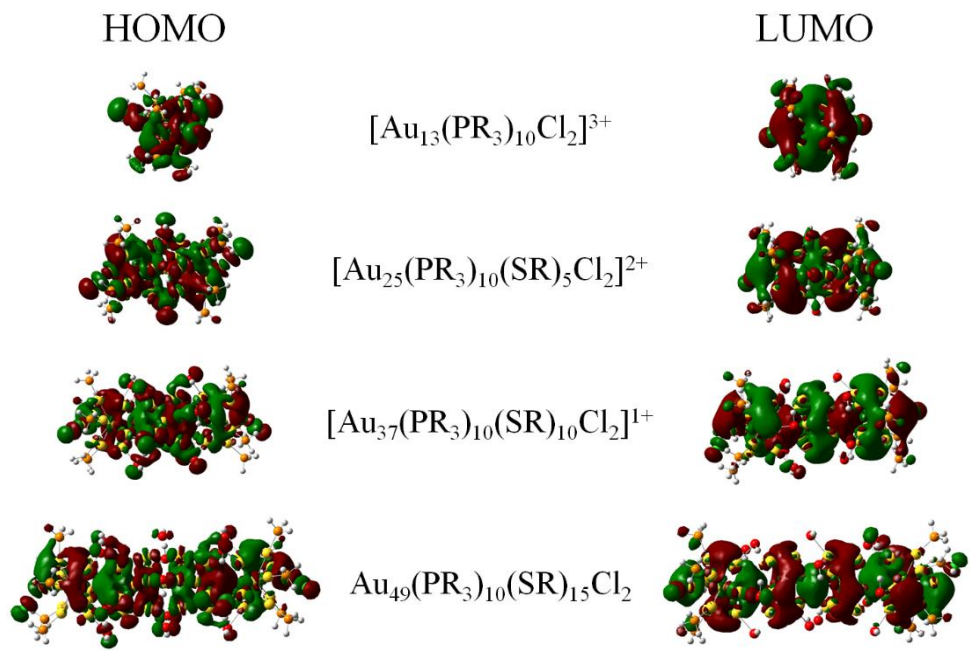

Figure S11. The HOMOs and LUMOs of [Au$_{13}$(PR$_3$)$_{10}$Cl$_2$]$^{3+}$, [Au$_{25}$(PR$_3$)$_{10}$(SR)$_5$Cl$_2$]$^{2+}$, [Au$_{37}$(PR$_3$)$_{10}$(SR)$_{10}$Cl$_2$]$^{1+}$, and Au$_{49}$(PR$_3$)$_{10}$(SR)$_{15}$Cl$_2$.

Table S1. 91 variants (I$_1$ – I$_{91}$) of valence states for the secondary block Au$_{13}$(8$e$). b, m, and t denote bottom flavor (1$e$), middle flavor (0.5$e$), and top flavor (0$e$), respectively.

| I$_1$ | [13b]$^{5+}$ | I$_2$ | [12b1m]$^{4.5+}$ | I$_3$ | [12b1t]$^{4+}$ |
|---|---|---|---|---|---|
| I$_4$ | [11b2m]$^{4+}$ | I$_5$ | [11b1m1t]$^{3.5+}$ | I$_6$ | [11b2t]$^{3+}$ |
| I$_7$ | [10b3m]$^{3.5+}$ | I$_8$ | [10b2m1t]$^{3+}$ | I$_9$ | [10b1m2t]$^{2.5+}$ |
| I$_{10}$ | [10b3t]$^{2+}$ | I$_{11}$ | [9b4m]$^{3+}$ | I$_{12}$ | [9b3m1t]$^{2.5+}$ |
| I$_{13}$ | [9b2m2t]$^{2+}$ | I$_{14}$ | [9b1m3t]$^{1.5+}$ | I$_{15}$ | [9b4t]$^{1+}$ |
| I$_{16}$ | [8b5m]$^{2.5+}$ | I$_{17}$ | [8b4m1t]$^{2+}$ | I$_{18}$ | [8b3m2t]$^{1.5+}$ |
| I$_{19}$ | [8b2m3t]$^{1+}$ | I$_{20}$ | [8b1m4t]$^{0.5+}$ | I$_{21}$ | [8b5t]$^{0}$ |
| I$_{22}$ | [7b6m]$^{2+}$ | I$_{23}$ | [7b5m1t]$^{1.5+}$ | I$_{24}$ | [7b4m2t]$^{1+}$ |
| I$_{25}$ | [7b3m3t]$^{0.5+}$ | I$_{26}$ | [7b2m4t]$^{0}$ | I$_{27}$ | [7b1m5t]$^{0.5-}$ |
| I$_{28}$ | [7b6t]$^{1-}$ | I$_{29}$ | [6b7m]$^{1.5+}$ | I$_{30}$ | [6b6m1t]$^{1+}$ |
| I$_{31}$ | [6b5m2t]$^{0.5+}$ | I$_{32}$ | [6b4m3t]$^{0}$ | I$_{33}$ | [6b3m4t]$^{0.5-}$ |
| I$_{34}$ | [6b2m5t]$^{1-}$ | I$_{35}$ | [6b1m6t]$^{1.5-}$ | I$_{36}$ | [6b7t]$^{2-}$ |
| I$_{37}$ | [5b8m]$^{1+}$ | I$_{38}$ | [5b7m1t]$^{0.5+}$ | I$_{39}$ | [5b6m2t]$^{0}$ |
| I$_{40}$ | [5b5m3t]$^{0.5-}$ | I$_{41}$ | [5b4m4t]$^{1-}$ | I$_{42}$ | [5b3m5t]$^{1.5-}$ |
| I$_{43}$ | [5b2m6t]$^{2-}$ | I$_{44}$ | [5b1m7t]$^{2.5-}$ | I$_{45}$ | [5b8t]$^{3-}$ |
| I$_{46}$ | [4b9m]$^{0.5+}$ | I$_{47}$ | [4b8m1t]$^{0}$ | I$_{48}$ | [4b7m2t]$^{0.5-}$ |
| I$_{49}$ | [4b6m3t]$^{1-}$ | I$_{50}$ | [4b5m4t]$^{1.5-}$ | I$_{51}$ | [4b4m5t]$^{2-}$ |
| I$_{52}$ | [4b3m6t]$^{2.5-}$ | I$_{53}$ | [4b2m7t]$^{3-}$ | I$_{54}$ | [4b1m8t]$^{3.5-}$ |
| I$_{55}$ | [4b9t]$^{4-}$ | I$_{56}$ | [3b10m]$^{0}$ | I$_{57}$ | [3b9m1t]$^{0.5-}$ |
| I$_{58}$ | [3b8m2t]$^{1-}$ | I$_{59}$ | [3b7m3t]$^{1.5-}$ | I$_{60}$ | [3b6m4t]$^{2-}$ |
| I$_{61}$ | [3b5m5t]$^{2.5-}$ | I$_{62}$ | [3b4m6t]$^{3-}$ | I$_{63}$ | [3b3m7t]$^{3.5-}$ |
| I$_{64}$ | [3b2m8t]$^{4-}$ | I$_{65}$ | [3b1m9t]$^{4.5-}$ | I$_{66}$ | [3b10t]$^{5-}$ |

| $I_{67}$ | $[2b11m]^{0.5-}$ | $I_{68}$ | $[2b10m1t]^{1-}$ | $I_{69}$ | $[2b9m2t]^{1.5-}$ |
|---|---|---|---|---|---|
| $I_{70}$ | $[2b8m3t]^{2-}$ | $I_{71}$ | $[2b7m4t]^{2.5-}$ | $I_{72}$ | $[2b6m5t]^{3-}$ |
| $I_{73}$ | $[2b5m6t]^{3.5-}$ | $I_{74}$ | $[2b4m7t]^{4-}$ | $I_{75}$ | $[2b3m8t]^{4.5-}$ |
| $I_{76}$ | $[2b2m9t]^{5-}$ | $I_{77}$ | $[2b1m10t]^{5.5-}$ | $I_{78}$ | $[2b11t]^{6-}$ |
| $I_{79}$ | $[1b12m]^{1-}$ | $I_{80}$ | $[1b11m1t]^{1.5-}$ | $I_{81}$ | $[1b10m2t]^{2-}$ |
| $I_{82}$ | $[1b9m3t]^{2.5-}$ | $I_{83}$ | $[1b8m4t]^{3-}$ | $I_{84}$ | $[1b7m5t]^{3.5-}$ |
| $I_{85}$ | $[1b6m6t]^{4-}$ | $I_{86}$ | $[1b5m7t]^{4.5-}$ | $I_{87}$ | $[1b4m8t]^{5-}$ |
| $I_{88}$ | $[1b3m9t]^{5.5-}$ | $I_{89}$ | $[1b2m10t]^{6-}$ | $I_{90}$ | $[1b1m11t]^{6.5-}$ |
| $I_{91}$ | $[1b12t]^{7-}$ | | | | |